\documentclass[a4paper,11pt]{article}
\usepackage{pos}
\usepackage{physics}
\usepackage{lipsum}
\usepackage{graphicx}
\usepackage{hyperref}
\usepackage{subcaption}

\title{Diquark mass and quark-diquark potential by lattice QCD using an extended HAL QCD method with a static quark.}

\author*[a]{Kai-Wen Kelvin-Lee}
\author[a]{Noriyoshi Ishii}

\affiliation[a]{Research Center for Nuclear Physics (RCNP), University of Osaka,\\
  10-1 Mihogaoka, Ibaraki, Osaka 567-0047}

\emailAdd{kelvin@rcnp.osaka-u.ac.jp}
\emailAdd{ishiin@rcnp.osaka-u.ac.jp}

\abstract{We  will  calculate  the  diquark  mass  together  with  the
  quark-diquark  potential.
  We apply an  extended HAL QCD potential method to  a baryonic system
  made up from a static quark and a diquark.
Numerical calculations are performed by employing 2+1 flavor QCD gauge
configurations  generated by  CP-PACS  and JLQCD  Collaborations on  a
\(16^{3} \times 32\) lattice with \(a^{-1} \approx 1.6\) GeV.
To  improve the  statistical noise  in the  propagators of  the static
quark, the HYP smearing is employed on the gauge links.
Two-point  correlators  of  quark-diquark  baryonic  system  are  then
computed to obtain their ground-state  energies where various types of
diquarks  are considered  (eg:  scalar  diquark, axial-vector  diquark
etc).
We apply an extended HAL QCD method  on a baryonic system made up from
a scalar diquark  and a static quark to study  the scalar diquark mass
and the  quark-diquark potential.
In order to  determine the diquark mass self-consistently  in this HAL
QCD method, we demand that the  baryonic spectrum in the p-wave sector
obtained from  the two-point correlators  should be reproduced  by the
potential obtained from the baryonic system in the s-wave sector.
We obtain  the scalar diquark  mass of roughly (2/3)\(m_{N}\)  , i.e.,
twice the  naïve estimates of  a constituent quark mass  together with
the quark-diquark potential of Cornell type (Coulomb + linear).}

\FullConference{The 21st International Conference on Hadron Spectroscopy and Structure (HADRON2025)\\
 27 - 31 March, 2025\\
Osaka University, Japan\\}

\begin{document}
\maketitle

\section{Introduction}

Diquarks are hypothesized to be fundamental building blocks of hadrons
and  are  believed   to  play  a  significant  role   in  various  QCD
phenomena~\cite{Jaffe_2005}.
A   diquark   consists   of   two  quarks,   and   the   decomposition
\(\mathbf{3}_{c} \otimes \mathbf{3}_{c}  = \bar{\mathbf{3}}_{c} \oplus
\mathbf{6}_{c}\)   indicates  that   diquarks   necessarily  carry   a
non-neutral color charge.
Among  the  possible  diquark  configurations,  the  so-called  “good”
diquark—namely,  the scalar  diquark  with quantum  numbers \(J^{P}  =
0^+\), isospin  \(I=0\), and color  \(\bar{\mathbf{3}}_c\)—is expected
to be the most significant.
This  configuration  is considered  the  lightest  diquark, as  it  is
energetically favored  by both  the color-magnetic interaction  in the
quark         model          and         the         instanton-induced
interaction~\cite{tHooft_1976,Shuryak_1982,Schäfer_Shuryak_1998,Shuryak_Zahed_2004}.
At high baryon number densities, such scalar diquarks are predicted to
condense in the  QCD vacuum, leading to spontaneous  breaking of color
SU(3)  symmetry and  the  emergence  of a  new  phase  known as  color
superconductivity.

Despite  their theoretical  importance, experimental  investigation of
diquarks remains  extremely challenging due to  QCD color confinement,
which prevents the observation of isolated diquarks.
First-principles studies using lattice QCD are therefore essential for
probing diquark properties.
However, lattice QCD studies of diquarks are themselves complicated by
color  confinement, which  renders standard  techniques developed  for
color-singlet hadrons ineffective.
For example, the conventional approach of extracting hadron masses via
exponential fits  to temporal  two-point correlators is  not justified
for diquarks, as their correlators do not exhibit bound-state poles in
momentum space (see Sec.~60.6.3 of Ref.~\cite{Workman_2022}).

Nevertheless,   several  lattice   QCD  studies   have  attempted   to
investigate diquark  masses over  the years, and  they can  be broadly
categorized into the following three approaches:
\begin{enumerate}
    \item  Applying  single-exponential   fits  to  diquark  two-point
      correlators  in the  Landau gauge~\cite{Hess_1997,  Babich_2007,
        Bi_2016}. \label{item:naive_fit}

    \item Introducing  a static quark  to neutralize the  color charge
      and    evaluating   gauge-invariant    quark–diquark   two-point
      correlators        to         extract        diquark        mass
      differences~\cite{C_Alexandrou_2006,             A_Francis_2022,
        Orginos_2006, Green_2010}. \label{item:static_quark}

    \item  Interpreting  the   diquark  mass  as  a   parameter  in  a
      non-relativistic quark–diquark potential  model and employing an
      extended  HAL  QCD method  to  extract  both  the mass  and  the
      potential~\cite{watanabe_2021,               K_watanabe_2022,
        Nishioka_2025}. \label{item:halqcd}
\end{enumerate}

In category~\ref{item:naive_fit}, stable plateaus  are observed in the
Landau-gauge  correlators,  which  are   then  fitted  with  a  single
exponential to estimate the diquark mass.
However, the absence of a true  bound-state pole in the correlators is
not adequately addressed.
Category~\ref{item:static_quark}    maintains   color    SU(3)   gauge
invariance by introducing a  static quark and constructing baryon-like
systems composed of the static quark and a diquark.
Diquark  mass differences  are estimated  from the  energy differences
between such baryonic systems.
However, these estimates do not explicitly account for the interaction
energy  between the  diquark and  the static  quark, which  introduces
uncertainty in the extracted diquark mass differences.
(Besides  mass  differences,  this approach  also  yields  interesting
insights into the spatial extent of diquarks.)

In category~\ref{item:halqcd}, an  extended HAL QCD method  is used to
construct a quark–diquark model in which the diquark mass appears as a
parameter.
Here,  \(\Lambda_c\) and  \(\Sigma_c\)  baryons are  treated as  bound
states of a charm quark and a diquark.
This approach avoids the issue of missing bound-state poles in diquark
two-point correlators.
However, the  extracted diquark mass  is sensitive to the  input charm
quark mass, which introduces ambiguity.
It should  be noted  that the  determination of  the charm  quark mass
itself is also not unique.

In the present work, we improve upon the above approaches by employing
an extended  HAL QCD method, with  a static quark replacing  the charm
quark in the \(\Lambda_c\)-like system.
This modification  eliminates the ambiguity associated  with the charm
quark mass encountered in category~\ref{item:halqcd}.
Moreover, by  combining the static quark  with the HAL QCD  method, we
also address the limitations of category~\ref{item:static_quark}.
Specifically, the interaction energy between  the static quark and the
diquark  can be  explicitly separated  from  the total  energy of  the
baryonic system,  thereby resolving  the uncertainty in  the extracted
diquark mass differences.

In  this paper,  we present  our results  for the  mass of  the scalar
diquark as  well as the  quark–diquark potential, obtained  using this
improved framework.

\section{Formalism}

Since  diquarks  carry   color  charge  $\bar{\mathbf{3}}_{c}$,  their
two-point correlators do not exhibit  physical poles in momentum space
due to color confinement~\cite{Workman_2022}.
Consequently,  direct  extraction  of  diquark  masses  from  temporal
two-point correlators via a single-exponential fit should be avoided.
To  overcome   this  issue,  we   follow  the  same  strategy   as  in
category~\ref{item:halqcd},  attempting  to  eliminate  the  ambiguity
associated with the charm quark mass by replacing the charm quark with
a static quark of infinite mass.

We first  note that the  angular momentum  of our baryonic  states can
essentially be  specified by the  orbital angular momentum $L$  of the
diquark relative to the static quark.
This is justified for two reasons:
First, we restrict ourselves to scalar diquarks in this study.
Second,  the spin  of the  static  quark is  completely decoupled  and
factorized  from the  rest of  the baryonic  system, since  the static
quark corresponds to  the heavy quark limit of the  charm quark (Heavy
Quark Spin Symmetry).
Accordingly,  the baryonic  state  can be  denoted as  $|B(L)\rangle$,
where $L$ is the orbital angular  momentum between the diquark and the
static quark.

We now consider the equal-time Nambu–Bethe–Salpeter (NBS) wavefunction
of the baryonic system composed of a static quark and a diquark:
\begin{equation}
  \psi_{L} (\mathbf{r})
  \equiv
  \bra{0}
  D_{c}(\mathbf{x})
  Q_{c}(\mathbf{y})
  \ket{B(L)}, 
  \quad
  \mathbf{r}
  \equiv
  \mathbf{x} - \mathbf{y},
\end{equation}
where  $Q_{c}(x)$ is  the static  quark field,  and $D_{c}(x)$  is the
composite scalar diquark field defined as
\begin{equation}
  D_{c}(x)
  \equiv
  \epsilon_{abc}
  u^{T}_{a}(x)
  C\gamma_{5}
  d_{b}(x),
\end{equation}
with $C \equiv i\gamma^2\gamma^0$ being the charge conjugation matrix.
We  require  that  this   NBS  wavefunction  satisfies  the  following
Schrödinger equation:
\begin{equation}
  \left( -\frac{\nabla^2}{2m_{D}} + \hat{V} \right)
  \psi_{L} (\mathbf{r})
  = (\varepsilon_{L} - m_{D})\, \psi_{L} (\mathbf{r}),
  \label{eq:schroedinger-eq}
\end{equation}
where $\varepsilon_{L}$  denotes the total relativistic  energy of the
baryonic state $\ket{B(L)}$.
Here, $m_{D}$ is  the diquark mass, which will be  determined later by
requiring consistency  between the baryonic masses  obtained from this
model and those extracted from two-point correlators.
The  quantity   \((\varepsilon_{L}  -  m_{D})\)  corresponds   to  the
``binding energy'' of the system.
The operator  $\hat{V}$ represents the quark–diquark  potential, which
is  derivatively expanded  as $\hat{V}  \simeq V_{0}(r)  + O(\nabla)$,
where $V_{0}(r)$ is the central (spin-independent) potential.
Due  to the  scalar nature  of the  $0^+$ diquark,  the spin–spin  and
tensor interactions are absent.
Moreover,  because  the  spin  of  the  static  quark  decouples,  the
spin–orbit interaction is also absent.

The equal-time  NBS wavefunction  is extracted from  the quark–diquark
four-point correlator.
For positive  time separations $t  > 0$, our  quark-diquark four-point
correlator is arranged as
\begin{align}
  C(\mathbf{r}, t; t_{\rm src}) &\equiv
  \frac{1}{V} \sum_{\mathbf{\Delta}} 
  \bra{0}
  D_c(\mathbf{r} + \mathbf{\Delta}, t) Q_c(\mathbf{\Delta}, t) \cdot \mathcal{J}^\dagger(t_{\rm src})
  \ket{0} \\
  &= \frac{1}{V} \sum_{\mathbf{\Delta}} \sum_{n} 
  \bra{0} D_c(\mathbf{x} + \mathbf{\Delta}, t) Q_c(\mathbf{\Delta}, t) \ket{n} 
  \bra{n} \mathcal{J}^\dagger(t_{\rm src}) \ket{0} 
  \cdot e^{-E_{n}(t - t_{\rm src})},
\end{align}
where $V$  is the spatial  volume, and the sum  over $\mathbf{\Delta}$
projects to zero spatial momentum.
The  states   $\ket{n}$  and   energies  $E_{n}$  denote   the  $n$-th
eigenstates and eigenvalues of the QCD Hamiltonian, respectively.
$\mathcal{J}(t_{\rm  src})$ denotes the source operator.
To enhance the overlap with the  ground state, we examine the wall and
the exponentially smeared sources with several smearing size.

To  specify   the  source  explicitly,  we   define  $q_a(f,t)  \equiv
\sum_{\mathbf{x}} q_a(\mathbf{x},t)  f(\mathbf{x})$ for $q =  u, d, Q$
with $f(\mathbf{x}): \mathbb{R}^3 \to \mathbb{C}$.
Then, $D_c(f,t) \equiv  \epsilon_{abc} u_a^T(f,t) C\gamma_5 d_b(f,t)$,
and the source operator is
\begin{equation}
  \mathcal{J}(t)
  \equiv
  D_{c'}(f,t) Q_{c'}(f,t),
\end{equation}
where  we choose  $f(\mathbf{x}) \equiv  1$  for the  wall source  and
$f(\mathbf{x}) \equiv \exp(-|\mathbf{x}|/b)$ with $b  = a, 2a, 3a$ for
smearing sources, where $a$ is the lattice spacing.

In the large-$t$ limit, the  four-point correlator is dominated by the
ground state,  and the S-wave NBS  wavefunction $\psi_{S}(\mathbf{r})$
is obtained as
\begin{equation}
  \psi_{S} (\mathbf{r})
  \propto
  C(\mathbf{r}, t; t_{\rm src})
  \quad
  \text{for large } t.
\end{equation}
Since we use wall and smeared sources, the extracted NBS wavefunctions
correspond to the S-wave.

We  define the  ``prepotential'' $\widetilde{V}_{0}(\mathbf{r})$  from
the NBS wavefunction as
\begin{equation}
  \widetilde{V}_{0}(\mathbf{r})
  \equiv
  \frac{\nabla^{2} \psi_{S}(\mathbf{r})}{\psi_{S}(\mathbf{r})}.
  \label{eq:prepotential}
\end{equation}
Applying  the Schrödinger  equation~\eqref{eq:schroedinger-eq} to  the
S-wave, we obtain
\begin{equation}
  \widetilde{V}_{0}(\mathbf{r})
  =
  2m_{D}
  \left[
    V_{0}(\mathbf{r}) + m_{D} - \varepsilon_{S}
    \right],
\end{equation}
where $\varepsilon_{S} \equiv \varepsilon_{L = S}$.

Using the  prepotential, we rewrite  the Schrödinger equation  for the
P-wave as the following ``pre-Schrödinger'' equation:
\begin{equation}
  \left(
  -\nabla^{2}
  + \widetilde{V}_{0}(r)
  \right)
  \psi_{P}(\mathbf{r})
  =
  \Delta \widetilde{E} \,
  \psi_{P}(\mathbf{r}),
  \label{eq:pre-schroedinger-eq}
\end{equation}
where  $\varepsilon_{P}  \equiv  \varepsilon_{L   =  P}$  and  $\Delta
\widetilde{E} \equiv 2m_D(\varepsilon_P - \varepsilon_S)$.

We determine the  diquark mass $m_D$ by  requiring consistency between
the baryonic masses obtained from  the two-point correlators and those
from the Schrödinger equation.
This is  done by solving Eq.~\eqref{eq:pre-schroedinger-eq}  using the
prepotential  $\widetilde{V}_0(\mathbf{r})$ in  the  P-wave sector  to
extract $\Delta \widetilde{E}$.
Then, $m_D$ is determined by
\begin{equation}
  m_{D}
  =
  \frac{\Delta \widetilde{E}}{2(\varepsilon_P - \varepsilon_S)},
  \label{eq:diquark-mass}
\end{equation}
using the values of $\varepsilon_P$ and $\varepsilon_S$ extracted from
two-point correlators.
Once $m_D$  is obtained, the quark–diquark  potential is reconstructed
as
\begin{equation}
  V_{0}(r)
  =
  \frac{1}{2m_{D}}
  \widetilde{V}_{0}(r)
  + m_{D}
  - \varepsilon_{S}.
  \label{eq:central-potential}
\end{equation}


\section{Numerical Results and Discussion}

\subsection{Lattice QCD Setup}

We perform our lattice QCD calculations  using the $2+1$ flavor ($ud +
s$)   gauge   configurations   generated    by   CP-PACS   and   JLQCD
Collaborations~\cite{T_Ishikawa_2008}.
This ensemble  consists of  454 configurations on  a $16^3  \times 32$
lattice, generated  with the renormalization-group  improved (Iwasaki)
gauge action at $\beta = 6/g^2 = 1.83$,
combined  with a  nonperturbatively  \(O(a)\)-improved (clover)  quark
action   at  $\kappa_{ud}   =   \kappa_{s}  =   0.13710$  (the   SU(3)
flavor-symmetric point) using $C_{\rm SW} = 1.761$.
This   setup  corresponds   to  a   lattice  spacing   of  $a   \simeq
0.1209(16)\  \mathrm{fm}$  ($a^{-1}   \approx  1.632\  \mathrm{GeV}$),
spatial  extent  $La \approx  1.934\  \mathrm{fm}$,  pion mass  $m_\pi
\simeq   1014\   \mathrm{MeV}$,   and   nucleon   mass   $m_N   \simeq
2026\ \mathrm{MeV}$~\cite{Inoue_2010}.

Before calculating  the two-point and four-point  correlators, Coulomb
gauge fixing is applied.
Hypercubic  (HYP)   smearing~\cite{Hasenfratz_Knechtli_2001}  is  then
applied to the  gauge links to construct the  static quark propagators
(Wilson lines), thereby reducing statistical fluctuations.
We use  both wall  sources and exponentially  smeared sources  for the
light-quark and static-quark propagators.
To  further  suppress  statistical  noise,  we  exploit  translational
symmetry in the temporal direction by averaging over 16 source points:
$t_{\rm src}/a = 0, 2, 4, \dots, 30$.
Rotational  symmetry  (cubic  group) and  combined  time-reversal  and
charge-conjugation symmetries are also used to reduce noise.
Statistical errors are estimated using the jackknife method with a bin
size of 10 configurations.

\subsection{Effective Mass Differences} \label{subsec:effective_mass_difference}

We consider  two-point correlators of  baryonic systems composed  of a
diquark and a static quark:
\begin{equation}
  G_{\Gamma}(t)
  \equiv
  \left\langle
  J_{\Gamma}(\mathbf{x},t)
  J^\dagger_{\Gamma}(\mathbf{x},0)
  \right\rangle,
\end{equation}
where  $J_{\Gamma}(x) \equiv  \epsilon_{abc}  \left(u^T_a(x) C  \Gamma
d_b(x)\right)  Q_c(x)$  denotes  the interpolating  operator  for  the
baryonic   state,    and   $\Gamma   =   1,    \gamma_\mu,   \gamma_5,
\gamma_5\gamma_\mu, \sigma_{\mu\nu}$.

The static quark propagator is replaced by a temporal Wilson line:
\begin{equation}
  S_{\rm stat}(\mathbf{x}_2,t_2; \mathbf{x}_1,t_1)
  =
  \delta^3(\mathbf{x}_2 - \mathbf{x}_1)
  \left( \frac{1 + \gamma_0}{2} \right)
  \left[
    \prod_{t=t_1}^{t_2-a} U_4(\mathbf{x}_1,t)
  \right]^\dagger
\end{equation}
for $t_2 >  t_1$, where the factor  $\exp\left(-m_Q(t_2 - t_1)\right)$
is dropped with $m_Q = \infty$ being the mass of the static quark.
The effective mass is defined in the standard way:
\begin{equation}
  m_{C\Gamma}(t)
  \equiv
  a^{-1}
  \log\left[
    G_{\Gamma}(t) / G_{\Gamma}(t+a)
    \right].
\end{equation}
Following Ref.~\cite{C_Alexandrou_2006}, we  define the effective mass
difference as
\begin{equation}
  \Delta m_{C\Gamma}(t)
  \equiv
  m_{C\Gamma}(t) - m_{C\gamma_5}(t),
\end{equation}
which is  the effective  mass difference  between the  baryonic states
with diquark channels $C\Gamma$ and $C\gamma_5$.
$\Delta m_{C\Gamma}(t)$ is a gauge invariant quantity, which can serve
as a  good estimate  of the  mass difference  between the  two diquark
channels according to the additive diquark model of heavy baryons (see
Sect.  8.4 of Ref.~\cite{Vogl_1991}).

\begin{figure}[htbp]
  \centering
  \includegraphics[width=0.6\textwidth]{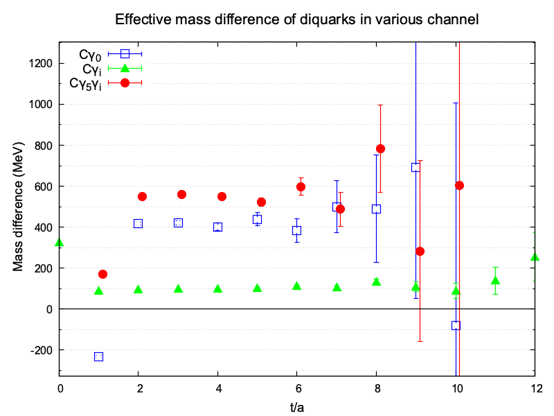}
  \caption{Effective  mass  differences   \(\Delta  m_{C\Gamma}\)  for
    baryonic  states   corresponding  to  various   diquark  channels,
    computed from wall-sourced two-point correlators.}
  \label{fig:mass_diff}
\end{figure}

Figure~\ref{fig:mass_diff}  shows   the  effective   mass  differences
obtained from wall-source propagators.
Our      results      are      consistent      with      those      in
Ref.~\cite{C_Alexandrou_2006}.
In the  positive parity  sector, the  ``bad'' axial-vector  diquark is
heavier than the ``good'' scalar diquark.
Negative-parity states are heavier than positive-parity states.
We also observe that, in both  parity sectors, vector diquarks tend to
be heavier than scalar diquarks.

We emphasize that our interpretation of the $1^{-}$ state differs from
that in Ref.~\cite{C_Alexandrou_2006}.
Whereas Ref.~\cite{C_Alexandrou_2006} treated this  state as a $1^{-}$
diquark in S-wave, we interpret it as a $0^{+}$ diquark in P-wave.
Our  interpretation is  supported by  phenomenological studies,  where
$\lambda$-mode    excitations    are    favored    over    $\rho$-mode
excitations~\cite{Nagahiro_2017}.
This  reinterpretation is  crucial  for applying  the  HAL QCD  method
presented later.

\subsection{NBS Wavefunction and Prepotential from the Four-Point Correlator}

\begin{figure}[htbp]
  \centering

  \begin{subfigure}{0.48\textwidth}
    \includegraphics[width=\linewidth]{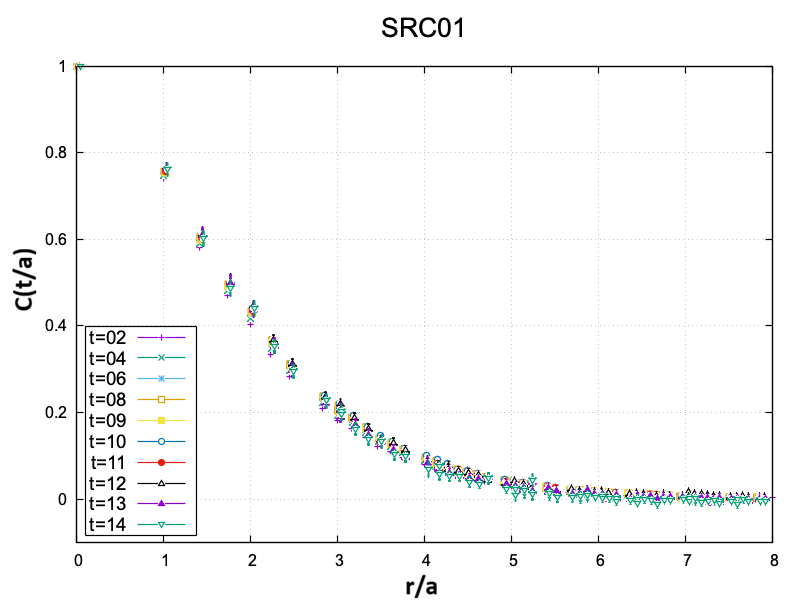}
  \end{subfigure}\hfill
  \begin{subfigure}{0.48\textwidth}
    \includegraphics[width=\linewidth]{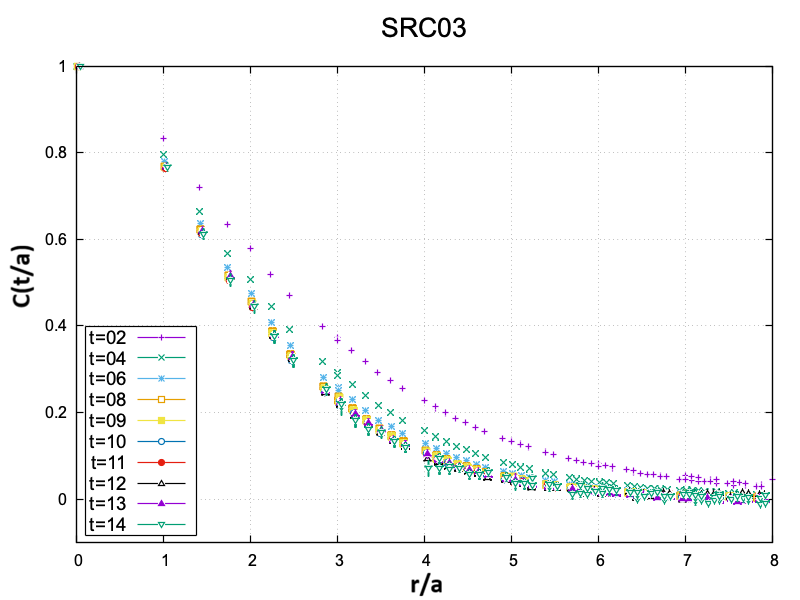}
  \end{subfigure}

  \medskip

  \begin{subfigure}{0.5\textwidth}
    \includegraphics[width=\linewidth]{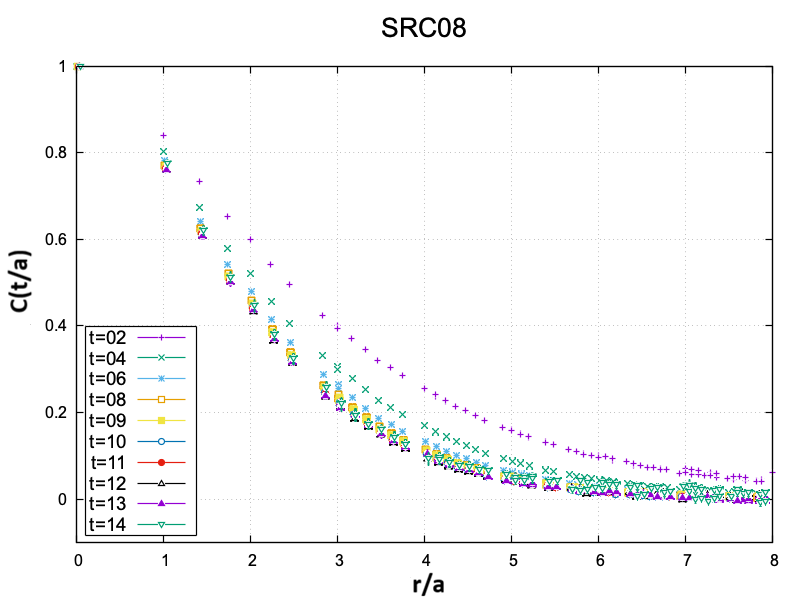}
  \end{subfigure}

  \caption{Convergence behavior of four-point correlators using three types of sources. SRC01 and SRC03 denote exponentially smeared sources with $f(r) = \exp(-r)$ and $\exp(-r/3)$, respectively (in lattice units). SRC08 denotes the wall source. Convergence rate: SRC01 > SRC03 > SRC08.} \label{fig:four_pt_func}
\end{figure}

The  normalized four-point  correlators  $\widetilde C(\mathbf{r},  t)
\equiv   C(\mathbf{r},  t)   /  C(\mathbf{0},   t)$  are   plotted  in
Fig.~\ref{fig:four_pt_func} as  a function  of $r =  |\mathbf{r}|$ for
fixed time slices $t = 2, 4, 6, \dots$.
We  use   three  types   of  sources:  SRC01   ($f(\mathbf{r})  \equiv
\exp(-|\mathbf{r}|/a)$), SRC03 ($\exp(-|\mathbf{r}|/(3a))$), and SRC08
(wall source).
For all  three, convergence is  roughly achieved at $t/a  \gtrsim 10$,
with SRC01 showing the fastest convergence.
We thus adopt SRC01 for subsequent analysis.

\subsection{Prepotential}

\begin{figure}[htbp]
  \centering
  \includegraphics[width=0.6\textwidth]{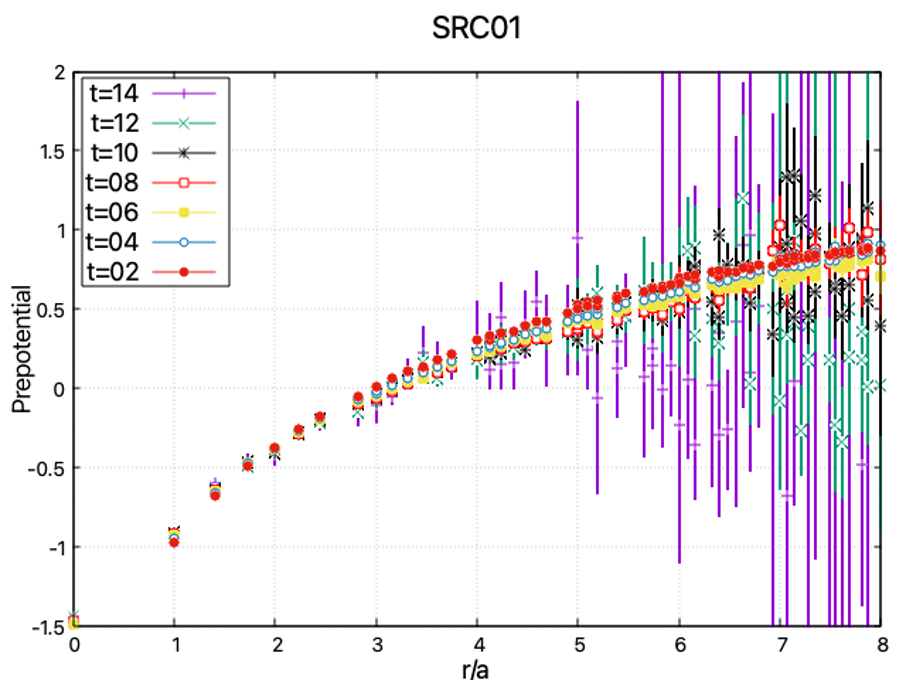}
  \caption{Prepotentials for $t/a = 2, 4, \dots, 14$ using SRC01.}
  \label{fig:prepotential}
\end{figure}

Figure~\ref{fig:prepotential}       shows      the       prepotentials
$\widetilde{V}_0(\mathbf{r})$  computed using  SRC01 for  various time
slices.
Convergence is observed at $t/a \gtrsim 10$.
We take the result at $t/a  = 10$ as the converged prepotential, which
we fit using a Cornell-type function:
\[
\widetilde{V}_0^{\text{fit}}(\mathbf{r})
=
-\frac{A}{r} + Br + v_0.
\]

\subsection{Scalar Diquark Mass $m_D$ and the Quark–Diquark Potential}

We  solve the  pre-Schrödinger equation~\eqref{eq:pre-schroedinger-eq}
in the P-wave sector using the fitted Cornell-type prepotential.
The bound-state solution is obtained  via the shooting method with the
Dormand–Prince          fifth-order          Runge–Kutta          (RK)
algorithm~\cite{Press_2007}.

Using   Eq.~\eqref{eq:diquark-mass}   with   the   extracted   $\Delta
\widetilde{E}$   and  the   effective  energies   $\varepsilon_P$  and
$\varepsilon_S$ from  the two-point correlators, we  obtain the scalar
diquark mass as
\begin{equation}
  m_D \simeq 1.241\ \mathrm{GeV},
\end{equation}
which  is  roughly  consistent  with   the  naive  estimate  based  on
constituent quark mass, i.e., $(2/3)m_N \simeq 1.35\ \mathrm{GeV}$.

\begin{figure}[htbp]
  \centering
  \includegraphics[width=0.6\textwidth]{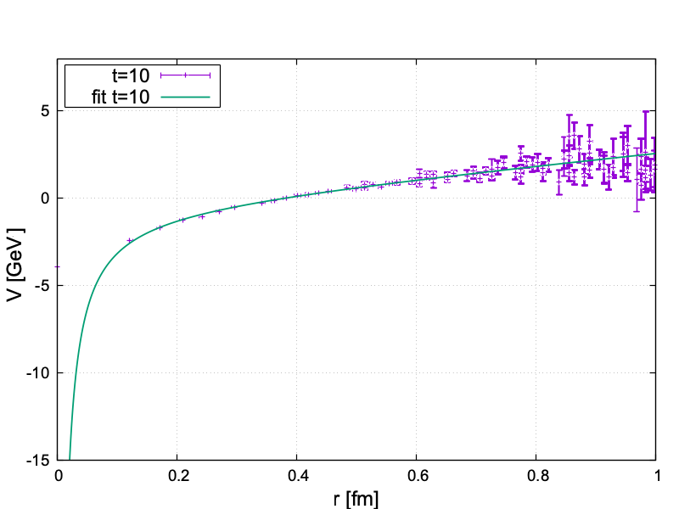}
  \caption{The quark–diquark potential.}
  \label{fig:potential}
\end{figure}

The     quark–diquark    potential     is    then     obtained    from
Eq.~\eqref{eq:central-potential},      and       is      shown      in
Fig.~\ref{fig:potential}.

Fitting this  potential with the  Cornell form $V_0(r)  = -A/r +  Br +
v_0$ yields:
\begin{align*}
  A        &= 0.121(4)\ \mathrm{GeV\cdot fm}, \\
  \sqrt{B} &= 510(5)\ \mathrm{MeV}, \\
  v_0      &= 0.141(15)\ \mathrm{GeV}.
\end{align*}
We note that HYP smearing affects short-distance behavior ($r \lesssim
2a$), which impacts the Coulomb coefficient $A$.
Our extracted  string tension $\sqrt{B}$  is slightly larger  than the
conventional value  from the  static quark–antiquark  potential ($\sim
440\ \mathrm{MeV}$)~\cite{SCHILLING_1993}.

\section{Conclusion and Future Outlook}

We have presented a lattice QCD  study of baryonic systems composed of
a diquark and a static quark, aiming to determine the diquark mass and
the quark–diquark potential.
Our calculations  are based on  $2+1$ flavor QCD  gauge configurations
generated by  CP-PACS and JLQCD  Collaborations on a $16^3  \times 32$
lattice.

Following  Ref.~\cite{C_Alexandrou_2006},  we computed  the  effective
mass  differences between  two baryonic  systems containing  different
diquarks.
These mass  differences provide  gauge-invariant estimates  of diquark
mass differences under  the additive diquark picture,  and our results
are consistent with those in Ref.~\cite{C_Alexandrou_2006}.

To overcome the issue of the  absence of a bound-state pole in diquark
two-point correlators, we treated the diquark mass as a mass parameter
within a quark–diquark model.
We constructed  this model by  applying an extended HAL  QCD potential
method to baryonic systems composed of a scalar (“good”) diquark and a
static quark.
Since  the static  quark  has  infinite mass,  our  method avoids  the
ambiguity associated with  the choice of charm quark  mass in previous
studies.
The diquark  mass was determined  by imposing a  consistency condition
between  the  baryonic mass  in  the  P-wave sector—obtained  via  the
Schrödinger equation—and  the baryonic  mass extracted  from two-point
correlators.
Our result  for the  scalar diquark  mass is  $m_D \simeq  1.241$ GeV,
which  is close  to $2m_N  /  3 \simeq  1.35$, i.e.,  twice the  naive
estimate of the constituent quark mass.

Importantly,  our  method allows  for  explicit  decomposition of  the
baryonic mass into  the sum of the diquark mass  and the quark–diquark
interaction energy.
While   Ref.~\cite{C_Alexandrou_2006}  interpreted   mass  differences
between baryonic  systems as  diquark mass differences,  our framework
enables a direct  and explicit determination of both  the diquark mass
and its mass differences from first principles.

We have also  extracted the quark–diquark potential,  which exhibits a
Cornell-like behavior  (Coulomb +  linear confinement), with  a string
tension  of  $\sqrt{\sigma} =  510(5)$  MeV—slightly  larger than  the
conventional value of $\sim 440$ MeV.

For future work,  we plan to apply variational  analysis with multiple
source operators to improve the precisions  of the diquark mass and the
quark-diquark potentials.
We will  extend the present  framework to the axial-vector  channel to
determine axial-vector diquark mass.
We  will investigate  the dependence  of  the diquark  masses and  the
quark-diquark  potentials differences  on the  quark mass  to make  an
extrapolation to the physical quark mass point.

\acknowledgments

Numerical   calculations  in   this   work  were   performed  on   the
supercomputer SQUID at D3 Center of Osaka University, supported by the
Research Center for Nuclear Physics (RCNP), Osaka University.
We thank the CP-PACS and JLQCD Collaborations as well as the JLDG/ILDG
for providing the $2+1$ flavor QCD gauge configurations.
We employed a modified version of  the lattice QCD library Bridge++ to
carry out our computations \cite{Ueda_2014}.
This work was supported by JSPS KAKENHI Grant Number JP21K03535.
We acknowledge  the support from  the Ministry of  Education, Culture,
Sports, Science  and Technology (MEXT)  of Japan through  the Japanese
Government Scholarship.

\bibliographystyle{unsrt}  
\bibliography{refs}        

\end{document}